\documentclass[a4paper]{article}

\usepackage{INTERSPEECH2020}

\usepackage[utf8]{inputenc}

\usepackage{url}

\title{Quantification of Transducer Misalignment in Ultrasound Tongue Imaging}
\name{Tamás Gábor Csapó$^{1,2}$, Kele Xu$^{3,4}$}
\address{
  $^1$Department of Telecommunications and Media Informatics, \\
	Budapest University of Technology and Economics, Budapest, Hungary \\
	$^2$MTA-ELTE Lendület Lingual Articulation Research Group, Budapest, Hungary \\
	$^3$National Key Lab of Parallel and Distributed Processing, \\ National University of Defense
Technology, Changsha, China \\
$^4$School of Computer, National University of Defense Technology, Changsha, China
 }
\email{csapot@tmit.bme.hu, kelele.xu@gmail.com}

\begin{document}

\maketitle
\begin{abstract}
In speech production research, different imaging modalities have been employed to obtain accurate information about the movement and shaping of the vocal tract. Ultrasound is an affordable and non-invasive imaging modality with relatively high temporal and spatial resolution to study the dynamic behavior of tongue during speech production. However, a long-standing problem for ultrasound tongue imaging is the transducer misalignment during longer data recording sessions. In this paper, we propose a simple, yet effective, misalignment quantification approach. The analysis employs MSE distance and two similarity measurement metrics to identify the relative displacement between the chin and the transducer. We visualize these measures as a function of the timestamp of the utterances. Extensive experiments are conducted on a Hungarian and Scottish English child dataset. The results suggest that large values of Mean Square Error (MSE) and small values of Structural Similarity Index (SSIM) and Complex Wavelet SSIM indicate corruptions or issues during the data recordings, which can either be caused by transducer misalignment or lack of gel.

  
\end{abstract}

\section{Introduction}

Ultrasound tongue imaging (UTI) is a technique suitable for the acquisition of articulatory data. Phonetic research has employed 2D ultrasound for a number of years for investigating tongue movements during speech \cite{Stone1983}. Stone summarized the typical methodology of investigating speech production using ultrasound~\cite{Stone2005a}. Usually, when the subject is speaking, the ultrasound transducer is placed below the chin, resulting in mid-sagittal images of the tongue movement. The typical result of 2D ultrasound recordings is a series of gray-scale images in which the tongue surface contour has a greater brightness than the surrounding tissue and air.
Compared to other articulatory acquisition methods (e.g.\ EMA, X-ray, XRMB, and vocal tract MRI), UTI has the advantage that the tongue surface is fully visible, and ultrasound can be recorded in a non-invasive way~\cite{Stone2005a,Csapo2017c,Ramanarayanan2018}. An ultrasound device is easy to handle and move, since it is small and light, and thus it is suitable for fieldworks, as well. Besides, it is a significantly less expensive piece of equipment than the above mentioned devices.

Clear applications for tongue ultrasound exist in linguistics, speech science, speech and swallowing therapy, and orthodontics \cite{Bressmann2005a, Rastadmehr2008, Hueber2011, Menard2012, Zharkova2013b}. Extracting tongue contours from ultrasound images is critical for later analyses, including comparison of tongue shapes, measuring parameters related to tongue curvature, addressing phonological questions related to articulation, and so on. In~\cite{Xu2016a} we proposed a novel automatic tongue contour tracking method which applies re-initialization as a potential solution for minimizing errors in tracking. In~\cite{Csapo2015a} we compared several other automatic tongue contour tracking methods which can be used for phonetic investigations, but found that all of them require manual corrections. Therefore, semi-automatic or interactive methods are preferred by linguists~\cite{Ghrenassia2014,Laporte2018}.

\begin{figure}[b]
\vspace{-4mm}
\centering
\includegraphics[trim=0.0cm 0.0cm 0.0cm 0.0cm, clip=true, height=33mm]{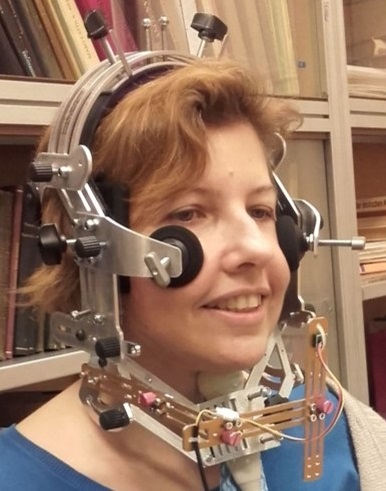}
\includegraphics[trim=0.0cm 0.0cm 0.0cm 0.0cm, clip=true, height=33mm]{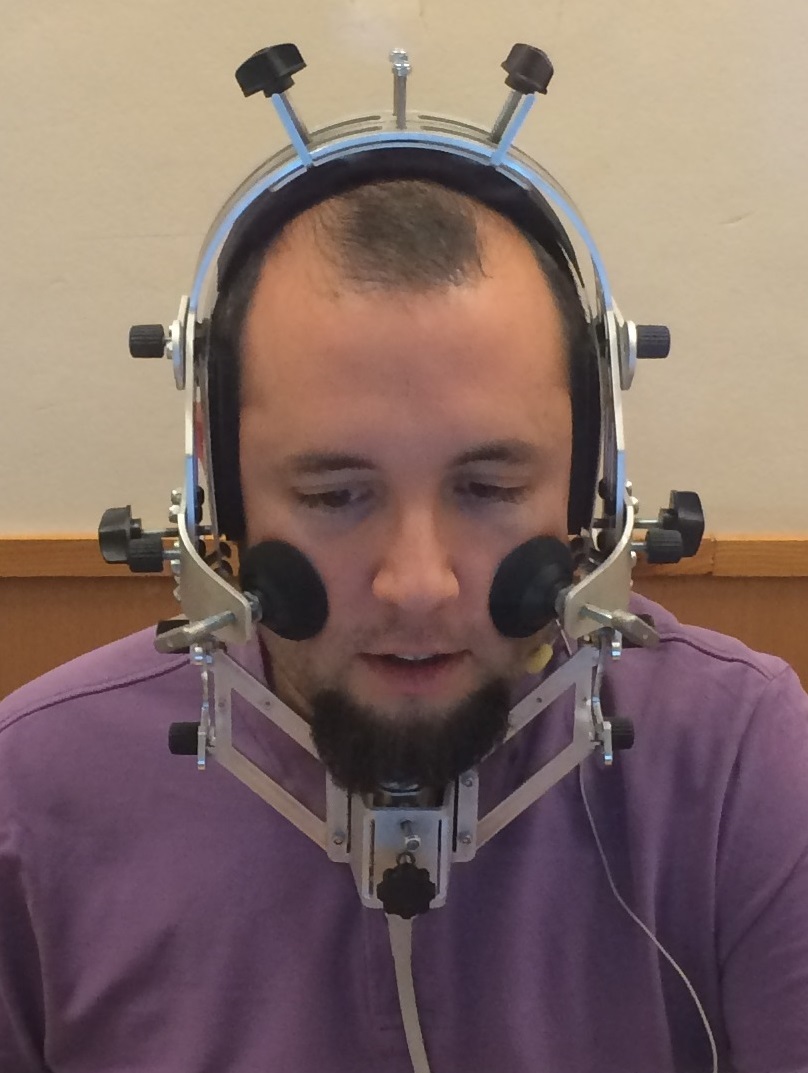}
\includegraphics[trim=0.0cm 0.0cm 0.0cm 0.0cm, clip=true, height=33mm]{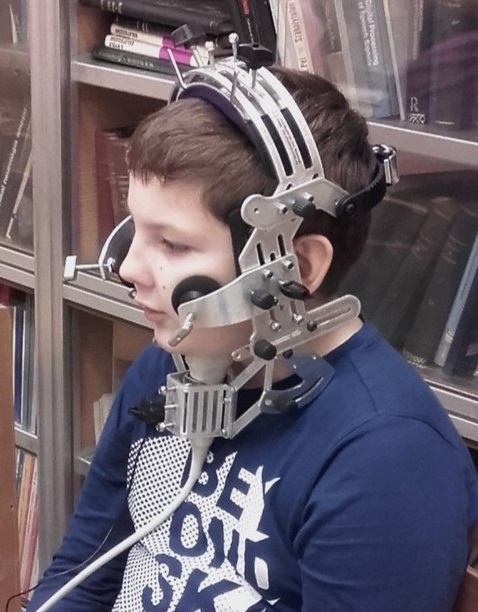}
\vspace{-1mm}
\caption{Recording of tongue movement with an ultrasound transducer fixing headset (Articulate Instruments Ltd.)~\cite{Csapo2017c,Marko2019b}.}
\vspace{-1mm}
\label{fig:ultrasound_headset}
\end{figure}

In order to fix head movement during the ultrasound recordings, various solutions have been proposed. The HATS system~\cite{Stone1995} aims to provide reliable tongue motion recordings by head immobilization and positioning the transducer in a known relationship to the head. Palatron~\cite{Mielke2005} is an algorithm to track the palate, thus could be used to align the ultrasound tongue images. The metal headset of Articulate Instruments Ltd.\ is a popular and well designed solution which was used in a number of studies (to mention a few, articulatory-to-acoustic mapping~\cite{Csapo2017c,Csapo2019}, Hungarian child recordings~\cite{Marko2019b,Graczi2020}, and UltraSuite~\cite{Eshky2018}). Fig.~\ref{fig:ultrasound_headset} shows a few samples. Hueber et al.\ proposed a set of accelerometers to track the position and orientation of the transducer, relative to the head~\cite{Hueber2011p}. Recently, UltraFit~\cite{Spreafico2018} is a lightweight headset to record ultrasound and EMA data.  

Despite these substantial efforts, it is a question whether the use of a headset itself is enough to ensure that the transducer is not moving during the recordings. Even when a transducer fixing system is used, large jaw movements during speech may cause the ultrasound transducer to move, resulting in misalignment. This way the recordings from the same session will not be directly comparable.
Although there exists methods for non-speech ultrasound transducer misalignment detection~\cite{Narayanan2014d,Bolsterlee2016}, they cannot be directly used in speech production research.

In this paper we analyze two datasets containing ultrasound tongue images, and show that transducer misalignment or other issues during recording could often happen. Besides, we quantify the amount of misalignment with three metrics.

\clearpage

\section{Methods}

\subsection{Ultrasound data}

We selected two databases containing synchronized speech and ultrasound tongue image recordings.

\vspace{-2mm}

\paragraph*{Hungarian children}

In the `Hungarian children' dataset~\cite{Marko2019b,Graczi2020}, two children, a girl and a boy read aloud nonsense words in five recording sessions within the course of 2 years. The girl was 7;5, 7;11, 8;5, 8;10 and 9;5 years old, the boy was 11;0, 11;6, 12;0, 12;5 and 13;0 years old at the time of the recording sessions. The tongue movement was recorded in midsagittal orientation using the ``Micro'' ultrasound system of Articulate Instruments Ltd. at 81.67 fps. The ultrasound transducer was fixed below the subjects' chin by the ultrasound stabilization headset designed for speech recordings (Articulate Instruments Ltd.), see Fig.~\ref{fig:ultrasound_headset}. The speech signal was recorded with a Beyerdynamic TG H56c tan omnidirectional condenser microphone. The ultrasound data and the audio signals were synchronized using the tools provided by Articulate Instruments Ltd. More details about the recording set-up can be found in~\cite{Csapo2017c}.

The linguistic material compiled for this small database contains trisyllabic nonsense words with $V_1 C_1 V_1 C_1 V_1$ structure, where all nine different vowel qualities of Hungarian are embedded within the context of nine voiceless obstruents: stops, fricatives and affricates. There is altogether 81 different sequences in this material, which were randomized and recorded in several repetitions in each children’s production (depending on the subject’s age and cooperation, the number of repetitions varies between 2 and 5). Before each repetition, swallowing (drinking water with a straw) was also recorded, for getting information about the palate. The order of randomization was the same across repetitions. The raw scanline data of the ultrasound was 64$\times$842 pixels.

\vspace{-2mm}

\paragraph*{UltraSuite}

We also employ the dataset from the publicly available UltraSuite repository~\cite{Eshky2018}, which contains the ultrasound tongue imaging data recorded for Scottish English children of two groups: typically developed children and the children with Speech Sound Disorders (SSD). Specifically, only the UXTD (typical developed) subset was used in our study. The raw scanline data of the ultrasound was 63$\times$412 pixels.





\subsection{Tongue contour data}

In an earlier study, we acquired manual tracings for a number of images in the `Hungarian children' dataset~\cite{Marko2019b,Graczi2020}. All Hungarian vowels in the middle of the target word were traced. Only the first two repetitions were included. Ultrasound images were extracted at the middle of the vowels, and tongue contours were manually traced using the APIL's web-based tracer tool (\url{https://github.com/myedibleenso/apil-web}).

The manually traced tongue contour data will be used to compare against the transducer misalignment measures.

\subsection{Transducer misalignment measures}

In order to quantify the amount of misalignment, we have chosen to compare consecutive recordings / utterances with each other. First, for a given speaker and given session, we go through all of the ultrasound recordings (utterances), and calculate the mean image (across time) of each utterance, including silences before and after. Next, we compare these mean images using the following image similarity measures. For a session, all of the consecutive utterances are compared with each other. If we have $n$ utterances in a session, than the result will be an $n \times n$ matrix containing the similarity measures, as in the case of DTW. We assume that if there is misalignment in the ultrasound transducer, than the matrix of measures will show this. Of course, the measures also indicate differences across the recordings caused by tongue movement (e.g.\ back /a/ vs.\ front /i/), but through longer recordings (containing neutral tongue position), it gets averaged out. Therefore, we expect that changes caused by ultrasound transducer misalignment will be clearly visible.

For the quantification, one error measure and two similarity metrics were chosen.

\vspace{-2mm}

\paragraph*{Mean Square Error (MSE)}

We first measured the MSE between the UTI pixels. \(y\) is one original mean image  and \(\hat{y}\) is another mean ultrasound image, while \(y_i\) and \(\hat{y}_i\) are their individual pixels. The calculations were done on the [0-255] grayscale pixel values.
MSE is an error measure, therefore the lower numbers indicate higher similarity across images.

\vspace{-2mm}

\paragraph*{Structural Similarity Index (SSIM)}

SSIM~\cite{Wang2004} (which was used earlier to compare ultrasound tongue images~\cite{Xu2016a}) measures three kinds of visual impact of changes in luminance \(l\), contrast \(c\) and structure \(s\) between two images:
\[
SSIM(y,\hat{y}) = [l(y,\hat{y})]^\alpha[c(y,\hat{y})]^\beta[s(y,\hat{y})]^\gamma
\]

In our experiment the SSIM index is calculated by 11$\times$11 circular-symmetric Gaussian weighting function, with standard deviation of 1.5~pixels.

\vspace{-2mm}

\paragraph*{Complex Wavelet Structural Similarity (CW-SSIM)}

CW-SSIM~\cite{Sampat2009} is an extension of the SSIM method to the complex wavelet domain, which is robust to small distortions:
\[
CW-SSIM(y,\hat{y}) = \frac{2\arrowvert\sum_{l=1}^{L}w_{y,l}w_{\hat{y},l}^{*}\arrowvert + K}{\sum_{l=1}^{L}|w_{y,l}|^2 + \sum_{l=1}^{L}|w_{\hat{y},l}|^2 + K}
\]

where \(w\) represents the complex wavelet coefficients of the two images. The \(^*\) indicates the complex conjugate of \(w\), and \(K\) is a small positive stabilizing constant~\cite{Xu2016a}.

In case of both the SSIM and CW-SSIM measures, the resulting range is between [0-1], and the higher value means more similar images (whereas zero is for the most diverse images).




\begin{figure*}
\centering
\includegraphics[trim=1.30cm 0.25cm 2.0cm 0.9cm, clip=true, width=0.24\textwidth]{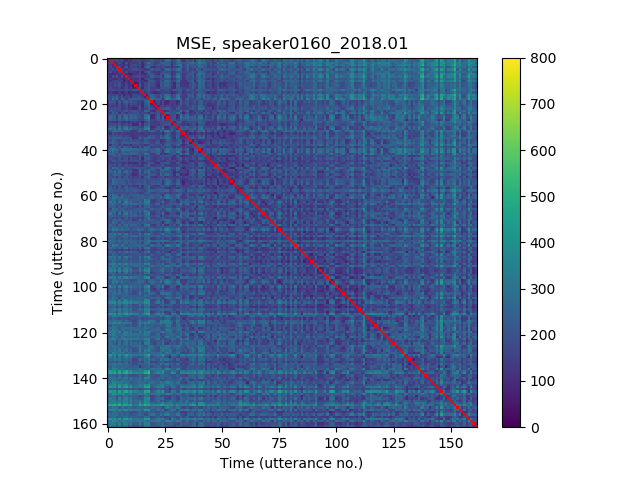}
\includegraphics[trim=1.30cm 0.25cm 2.0cm 0.9cm, clip=true, width=0.24\textwidth]{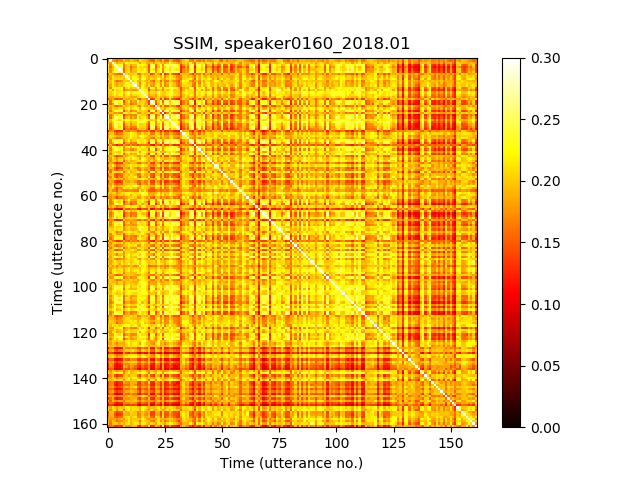}
\includegraphics[trim=1.30cm 0.25cm 2.0cm 0.9cm, clip=true, width=0.24\textwidth]{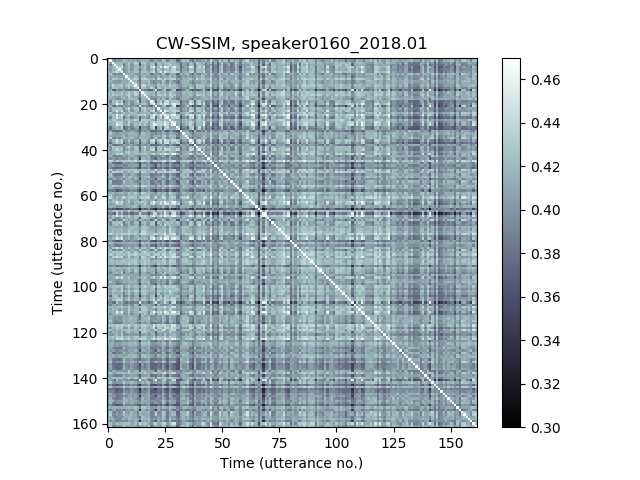}
\includegraphics[trim=-1.3cm -1.5cm -1.3cm 0.0cm, clip=true, width=0.22\textwidth]{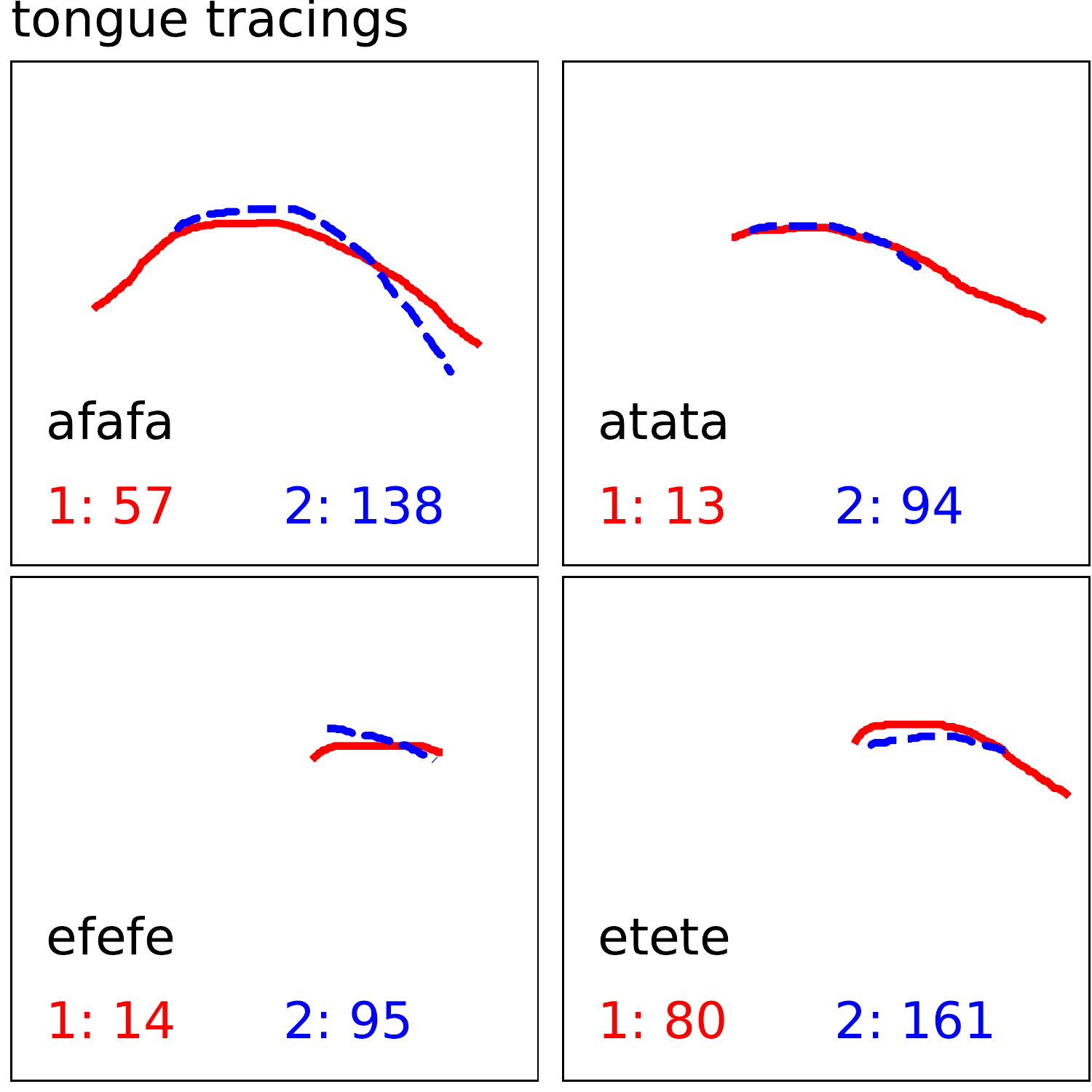}
\vspace{-2mm}
\caption{Sample for well aligned data across two repetitions, from the `Hungarian children' dataset. Repetition 1: utterances 1--81; repetition 2: utterances 82--162. MSE: lower values (blue colors) indicate smaller misalignment. SSIM: higher values (yellow color) indicate smaller misalignment. CW-SSIM: higher values (light gray) indicate smaller misalignment. The diagonals contain NaN values. In the tongue tracing figure (last column), 1: 57 denotes that the first repetition is utterance no.\ 57.}
\label{fig:UTI_HUN_aligned}
\vspace{-2mm}
\end{figure*}

\begin{figure*}
\centering
\includegraphics[trim=1.30cm 0.25cm 2.0cm 0.9cm, clip=true, width=0.24\textwidth]{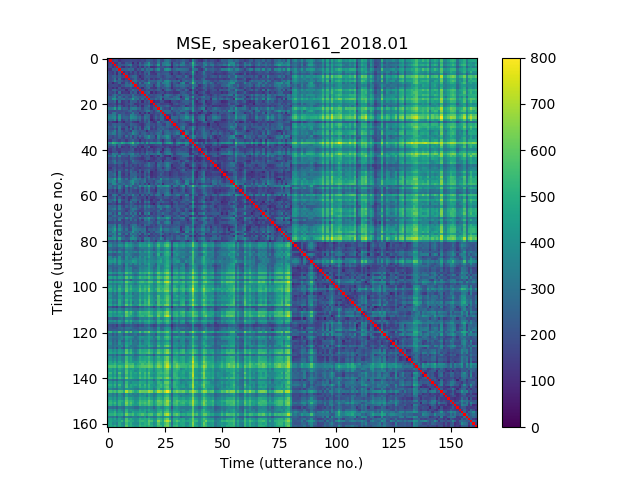}
\includegraphics[trim=1.30cm 0.25cm 2.0cm 0.9cm, clip=true, width=0.24\textwidth]{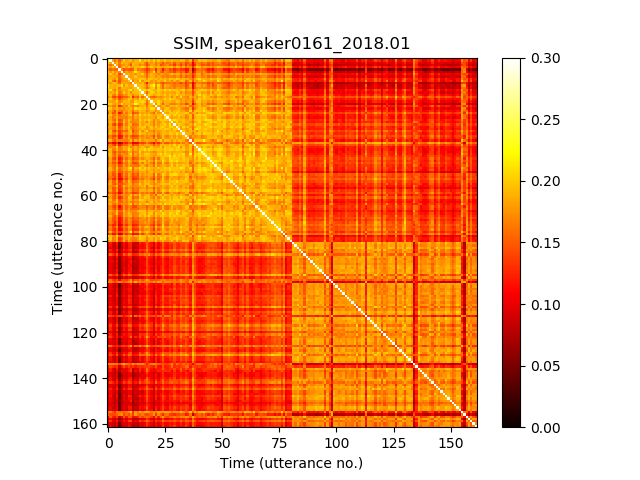}
\includegraphics[trim=1.30cm 0.25cm 2.0cm 0.9cm, clip=true, width=0.24\textwidth]{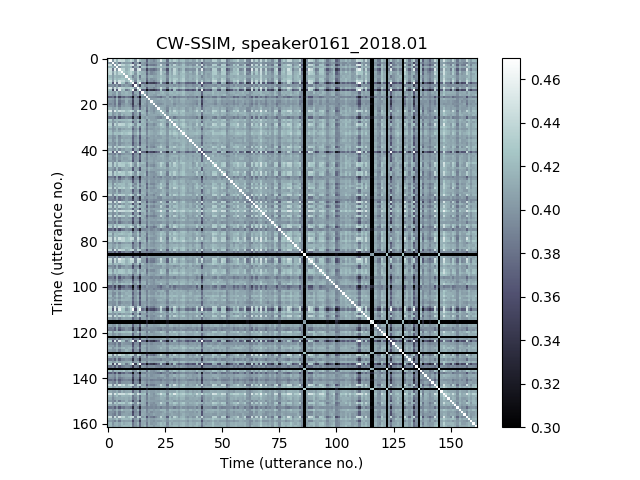}
\includegraphics[trim=-1.3cm -1.5cm -1.3cm 0.0cm, clip=true, width=0.22\textwidth]{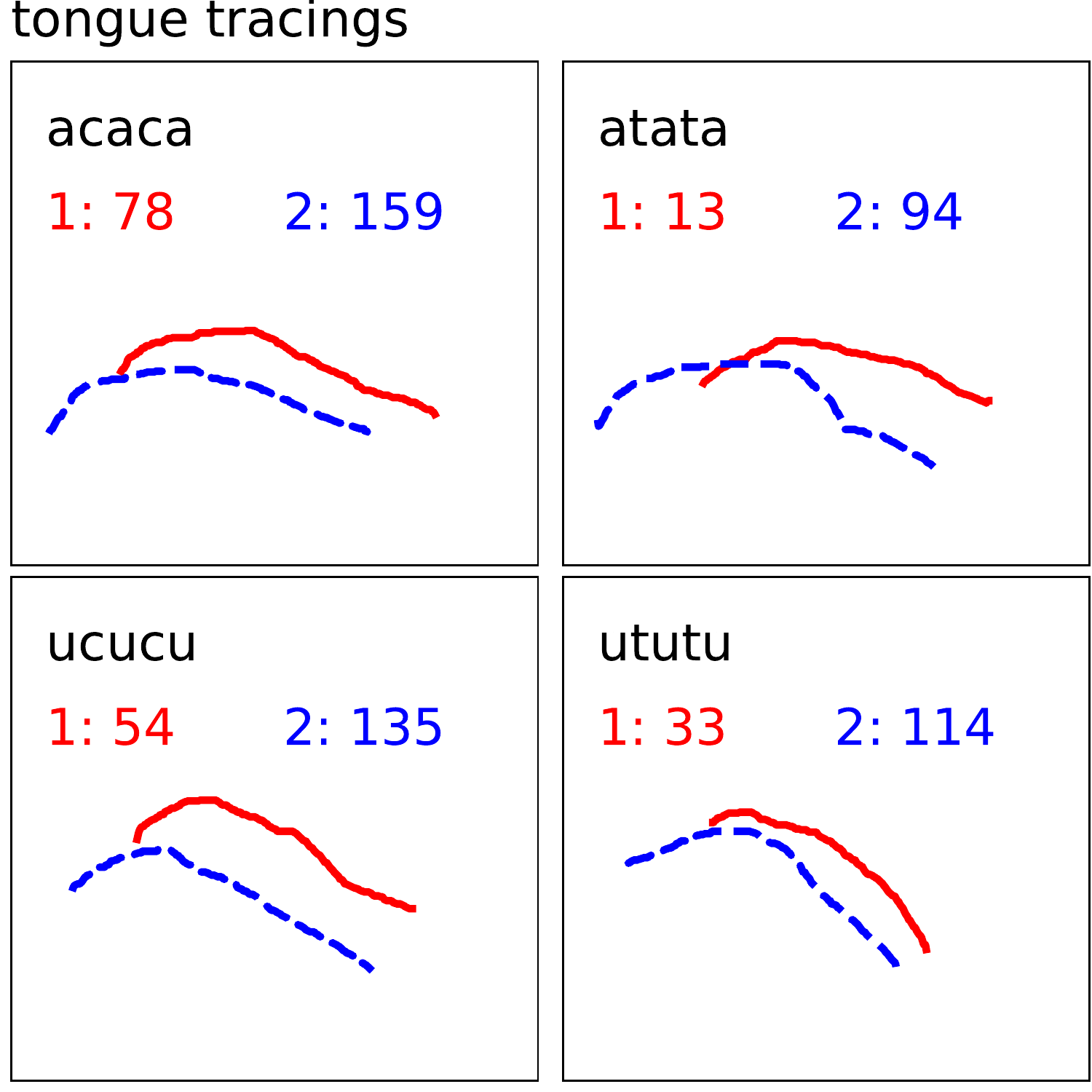}
\vspace{-2mm}
\caption{Sample for misalignment across two repetitions, from the `Hungarian children' dataset. Repetition 1: utterances 1--81; repetition 2: utterances 82--162.}
\label{fig:UTI_HUN_misaligned}
\vspace{8mm}
\end{figure*}

\begin{figure*}
\centering
\includegraphics[trim=1.30cm 0.25cm 2.0cm 0.9cm, clip=true, width=0.24\textwidth]{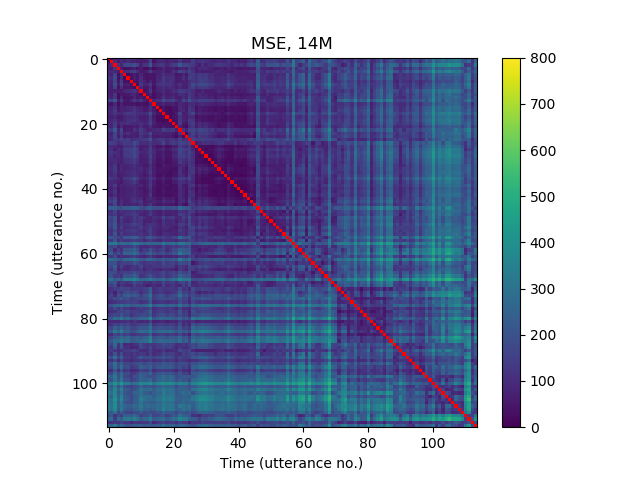}
\includegraphics[trim=1.30cm 0.25cm 2.0cm 0.9cm, clip=true, width=0.24\textwidth]{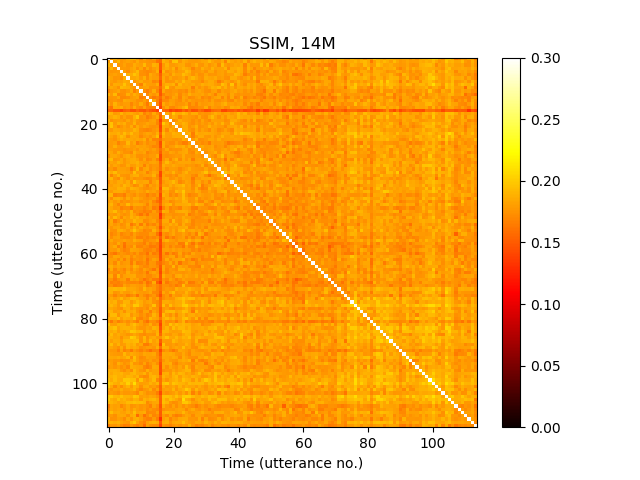}
\includegraphics[trim=1.30cm 0.25cm 2.0cm 0.9cm, clip=true, width=0.24\textwidth]{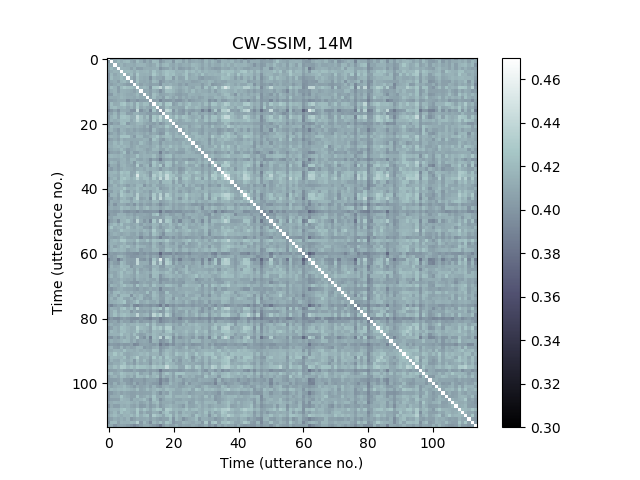}
\includegraphics[trim=-0.8cm 0.2cm -0.8cm 0.5cm, clip=true, width=0.24\textwidth]{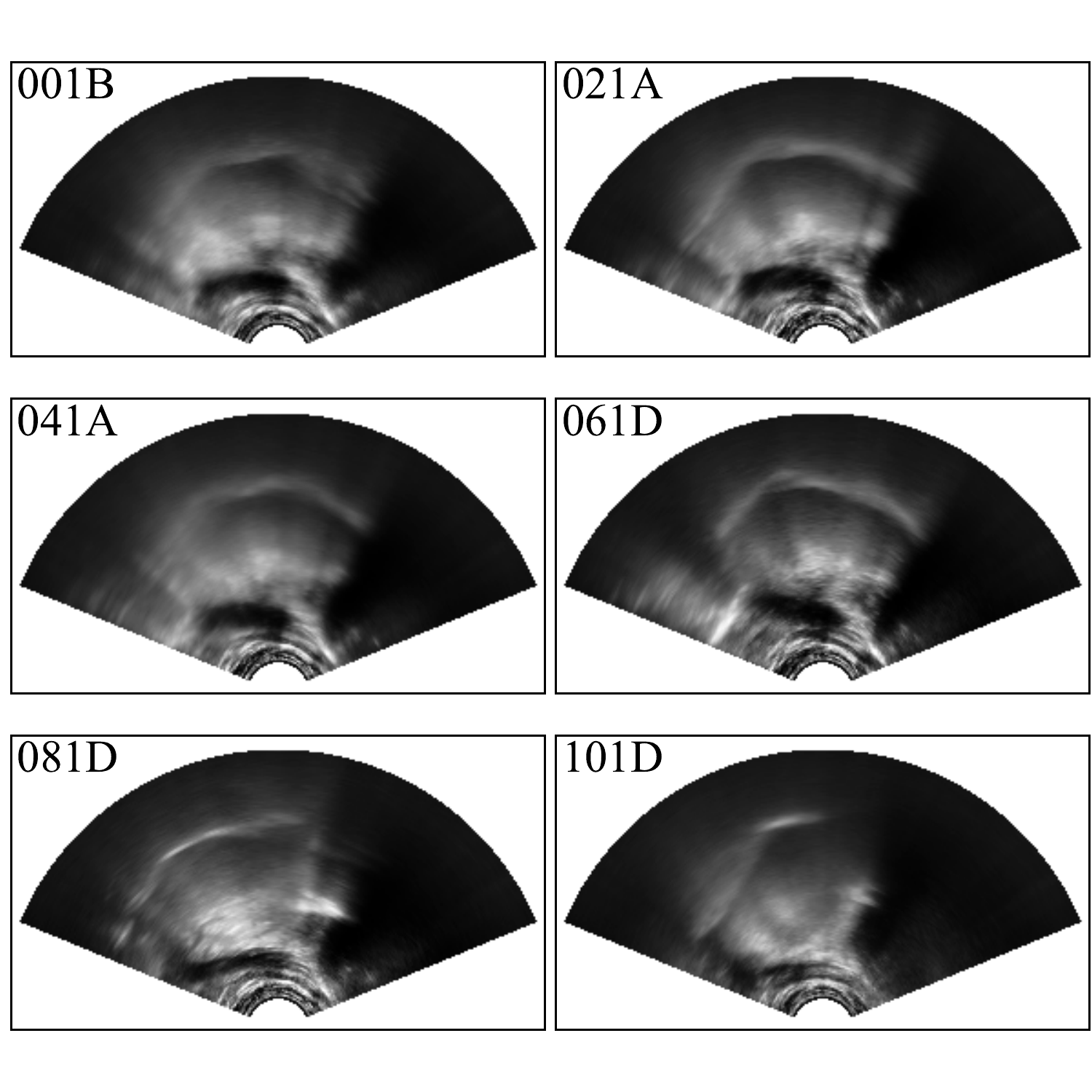}
\vspace{-2mm}
\caption{Sample for well aligned data, from the `UltraSuite' dataset. Columns 1--3: measures; column 4: ultrasound samples. The matrices for all UltraSuite / UXTD speakers can be found at \protect\url{https://github.com/BME-SmartLab/UTI-misalignment/}}
\vspace{-2mm}
\label{fig:UltraSuite_aligned}
\end{figure*}

\begin{figure*}
\centering
\includegraphics[trim=1.30cm 0.25cm 2.0cm 0.9cm, clip=true, width=0.24\textwidth]{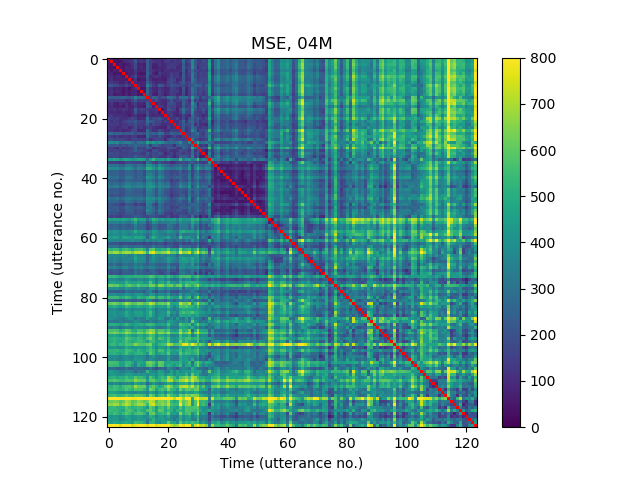}
\includegraphics[trim=1.30cm 0.25cm 2.0cm 0.9cm, clip=true, width=0.24\textwidth]{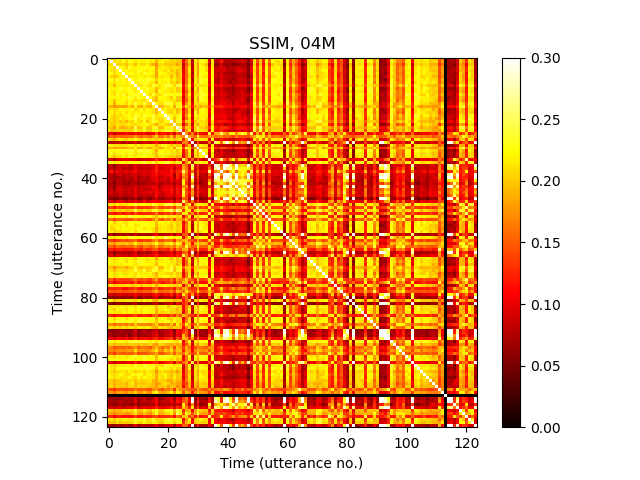}
\includegraphics[trim=1.30cm 0.25cm 2.0cm 0.9cm, clip=true, width=0.24\textwidth]{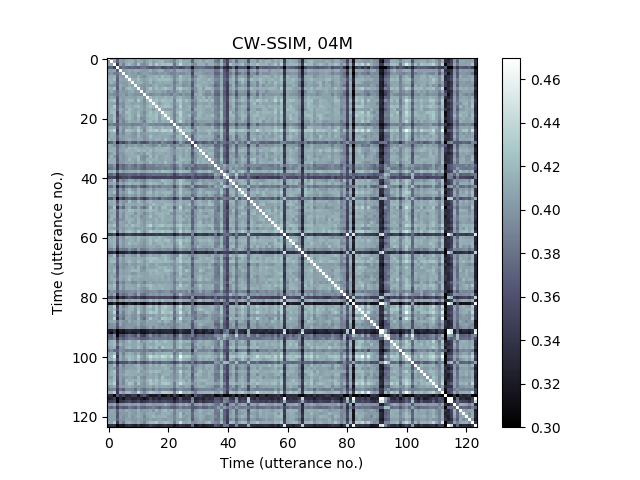}
\includegraphics[trim=-0.8cm 0.2cm -0.8cm 0.5cm, clip=true, width=0.24\textwidth]{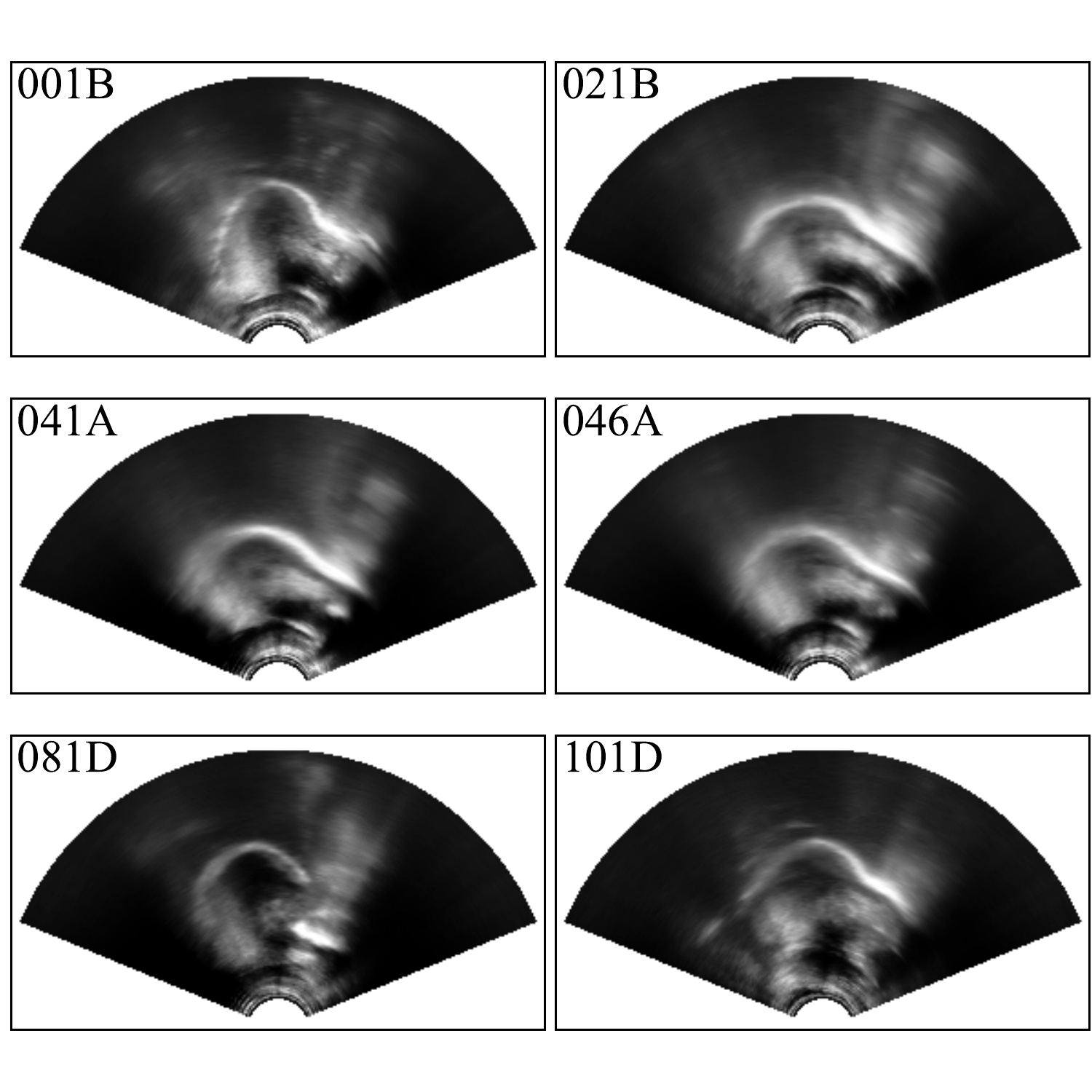}
\vspace{-2mm}
\caption{Sample for misaligned data, from the `UltraSuite' dataset.}
\label{fig:UltraSuite_misaligned}
\vspace{-2mm}
\end{figure*}

\begin{figure*}
\centering
\includegraphics[trim=1.30cm 0.25cm 2.0cm 0.9cm, clip=true, width=0.24\textwidth]{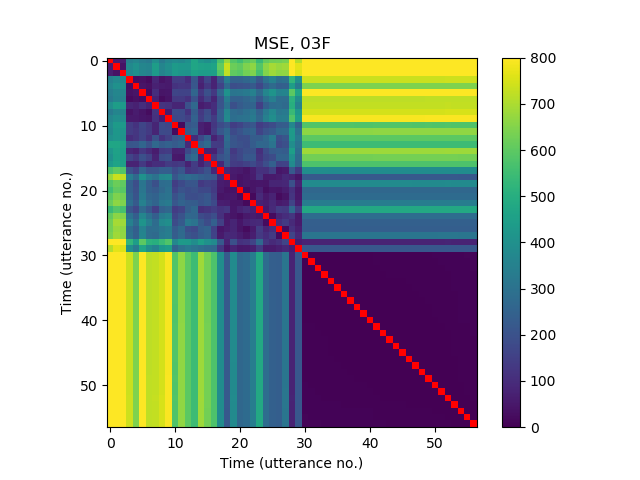}
\includegraphics[trim=1.30cm 0.25cm 2.0cm 0.9cm, clip=true, width=0.24\textwidth]{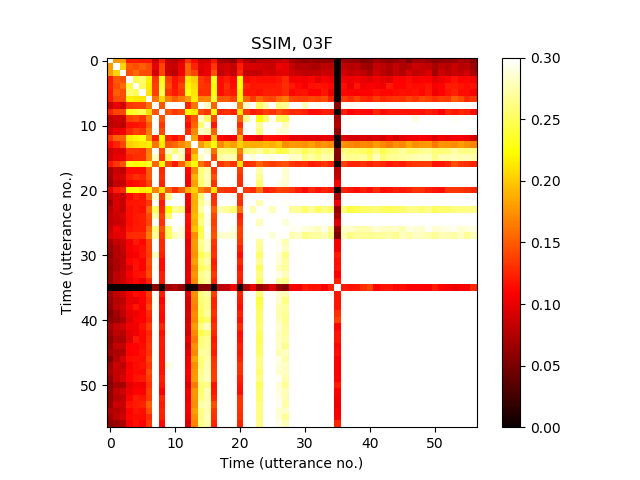}
\includegraphics[trim=1.30cm 0.25cm 2.0cm 0.9cm, clip=true, width=0.24\textwidth]{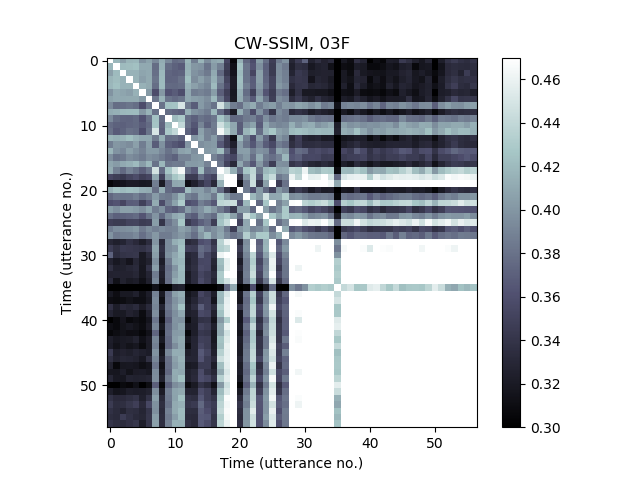}
\includegraphics[trim=-0.8cm 0.2cm -0.8cm 0.5cm, clip=true, width=0.24\textwidth]{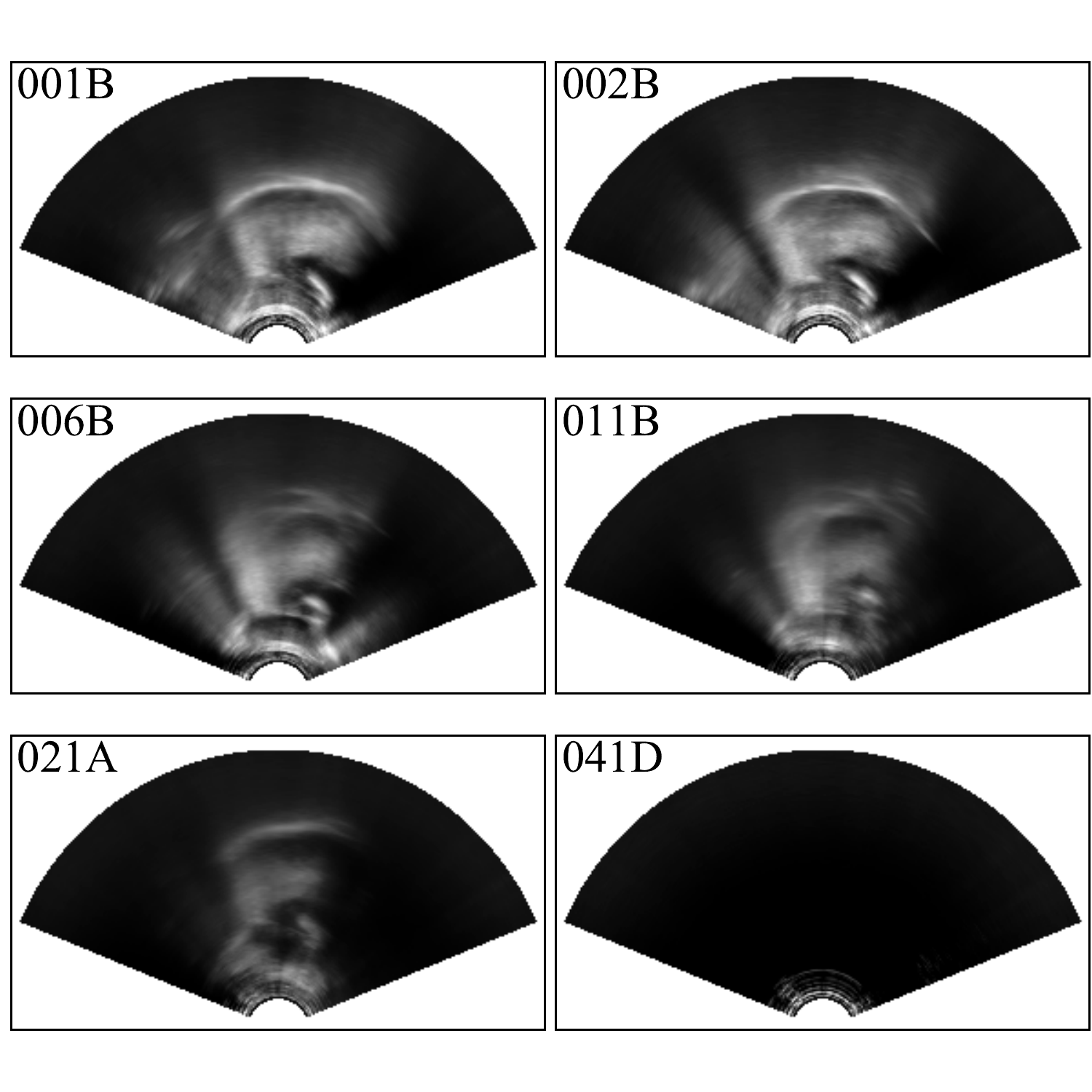}
\vspace{-2mm}
\caption{Sample for corrupted data, from the `UltraSuite' dataset.}
\label{fig:UltraSuite_corrupted}
\end{figure*}


\section{Results and discussion}

The results are demonstrated in Figures~\ref{fig:UTI_HUN_aligned}--\ref{fig:UltraSuite_corrupted}. The figures contain samples from both the Hungarian and English databases, with a few speakers hand-selected for visualization purposes. The objective distance (MSE) and similarity measures (SSIM and CW-SSIM) are shown for well aligned, misaligned and corrupted ultrasound utterance sequences.

\subsection{Results on the `Hungarian children' dataset with} 

Fig.~\ref{fig:UTI_HUN_aligned} shows the three measures (MSE, SSIM and CW-SSIM) in the first three columns, and several manual tracings in the last column (left: tongue root, right: tongue tip). We selected this sample to show the measures when the transducer did not move off within the recording session. In the MSE figure, all colors are blueish, indicating that MSE across most images is relatively small. SSIM shows large similarity for utterances no. 1--125, while there seems to be some difference between  utterances 1--125 vs.\ 126--162. CW-SSIM also presents this, but with weaker intensity differences. The manual tongue tracings in the last column of Fig.~\ref{fig:UTI_HUN_aligned} show that in terms of tongue contour, the two repetitions are similar; indicating that there was no (or only minimal) misalignment during the recording session.

For a sample containing clear misalignment, we have chosen the data visualized in Fig.~\ref{fig:UTI_HUN_misaligned}. According to both MSE and SSIM, utterances 1--81 are highly different from utterances 82--162. Meanwhile, differences within both utterances 1--81 and 82--162 are small. This might be caused by the fact that after each repetition (i.e., between utterances 81 and 82), the subject took a small break and was instructed to drink water for recording swallow. Most probably, the ultrasound headset went off during this break. In the last column, the manually traced tongue contours support the previous finding: the second repetition (blue line) is shifted upper and right compared to the first repetition (red dashed line). In our earlier study which was comparing the tongue contours~\cite{Marko2019b,Graczi2020}, we had the same observation, when measuring Nearest Neighbor Distance (NND)~\cite{Zharkova2011} between the tongue contours. However, NND is not suitable to quantify this type of shift, because it calculates the point-by-point minimal differences, and not the shift of the points.  


The means and standard deviations of the measures are shown in Table~\ref{tab:measures}. In terms of MSE, we can observe that both the mean and standard deviation is smaller for the well aligned data (speaker 0160 / 2018.01) than for the misaligned data (speaker 0161 / 2018.01). Also, the std.~dev.~of CW-SSIM is twice as large for the misaligned data than for the well aligned data.


\subsection{Results on the UltraSuite dataset}

For the English children dataset, in Fig.~\ref{fig:UltraSuite_aligned} we show first a sample with small differences across the recording session. According to MSE, there is no larger difference across utterances 1--47; and after this, the differences are also relatively small. SSIM shows the same tendency, except for utterance 18, which is clearly different from the rest. CW-SSIM does not indicate any issue. Some demonstration mean ultrasound images were converted to wedge orientation using ustools (\url{https://github.com/UltraSuite/ultrasuite-tools}). The mean images are similar across the whole recording session.

In Fig.~\ref{fig:UltraSuite_misaligned}, we show an example for larger within-session differences for the 04M speaker of the UltraSuite dataset. For utterances 1--32 and 34--53, MSE shows small difference. But across these two parts, and after utterance 54, all the MSE matrix indicates large differences across the utterances. Although we do not know the recording conditions for UltraSuite, for speaker 04M, but here it could happen that the transducer-fixing headset was not tight enough (or was not compatible with the small head of the child), and random transducer misalignment happened during the whole recording session. Another issue could have been that after several utterances, there remained little gel between the ultrasound transducer and skin, in the direction of the back and front parts of the tongue. The mean ultrasound images in the last columns also support these observations (e.g.\ utterances 041A and 046A are similar to each other, but clearly different from 081D and 101D).

Fig.~\ref{fig:UltraSuite_corrupted} presents another kind of corruption in the UltraSuite dataset, which happened for speaker 03F. Between utterances 3--18, and 19--28, the MSE is relatively small (whereas it is higher when comparing these two ranges). Starting from utterance 30, the MSE is extremely small. But in this case, this does not indicate well aligned transducer position. If we check the ultrasound images itself, we can see that the transducer went completely off (e.g.\ there was no more gel between the top of the transducer and the skin), and the tongue movement was not recorded between utterances 30--55. When checking SSIM and CW-SSIM, the same tendencies are visible. The mean images in the last column also show that in the last utterances (041D), the tongue surface is not visible because of the lack of gel. 

Table~\ref{tab:measures} summarizes the descriptive statistics for the three visualized speakers of the UltraSuite dataset. In terms of MSE, the mean is clearly higher for 04M (misaligned) than for 14M (well aligned). On the other hand, the mean of MSE is also high for 03F, because of the tongue movement not visible on 20 utterances. SSIM and CW-SSIM seem to be less reliable here.

\begin{table}
\caption{Descriptive statistics of the misalignment measures from selected speakers. Full data: \protect\url{https://github.com/BME-SmartLab/UTI-misalignment/ } .} \label{tab:measures}
\vspace{-2mm}
\centering
\begin{tabular}{l||c|c|c}
              & MSE & SSIM &  CW-SSIM \\
{\bf Speaker} & mean (std) & mean (std) &  mean (std) \\
\hline\hline
0160\_2018.01 & 219 ( 67) & 0.19 (0.03) & 0.41 (0.02) \\
0161\_2018.01 & 322 (128) & 0.15 (0.03) & 0.39 (0.04) \\
\hline
03F & 351 (313) & 0.28 (0.15) & 0.41 (0.07) \\
04M & 368 (156) & 0.16 (0.06) & 0.39 (0.03) \\
14M & 178 ( 91) & 0.18 (0.01) & 0.41 (0.01) \\

\end{tabular}
\vspace{-6mm}
\end{table}


\subsection{General discussion}

In order to fix the ultrasound transducer position during recordings, various approaches can be used (e.g.\ HATS~\cite{Stone1995}, Palatron~\cite{Mielke2005}, Articulate Instruments Ltd.\ headset~\cite{Eshky2018}, transducer orientation tracking~\cite{Hueber2011p}, or UltraFit~\cite{Spreafico2018}), but none of these methods are perfect and they cannot guarantee that tongue position or orientation will be the same in a longer recording session. If it is important that the repetitions are at the same position, other methods are suggested to be used besides ultrasound transducer position fixing. An example for this is the measurement of the occlusal plane with a biteplate and rotating / shifting the data to a reference coordinate system~\cite{Scobbie2011,Scobbie2012,Percival2020}. However, it requires significant amount of manual work, and according to our knowledge, until now there have been no methods yet for auto-rotating within longer ultrasound recording sessions.


\section{Conclusions}

In this paper we showed an attempt for the quantification of ultrasound transducer misalignment, on articulatory recordings of tongue movement. We used three measures (MSE, SSIM and CW-SSIM) for investigating changes in average ultrasound image sequences as a function of time. We hypothesized that large values of MSE and small similarity values (in terms of SSIM and CW-SSIM) indicate corruptions or issues during the data recordings, which can either be caused by transducer misalignment or lack of gel. Although we did not attempt to show a direct relationship between the quantified measures and amount of shift in tongue tracings, the results might be useful for phonetic research investigating tongue shapes and positions.

In the future we plan to develop automatic classification methods to warn during analysis of the tongue contours if the ultrasound transducer is clearly misaligned within a recording session; or give confidence intervals related to the reliability.

The code implementations are accessible at \url{https://github.com/BME-SmartLab/UTI-misalignment/}.

\section{Acknowledgements}

The first author was partly funded by the National Research, Development and Innovation Office of Hungary (FK 124584 and PD 127915 grants). We would like to thank the Ultrax2020 project for providing the UltraSuite articulatory database.

\clearpage

\bibliographystyle{IEEEtran}

\bibliography{ref_collection_csapot_nourl}

\end{document}